\documentclass[11pt,a4paper]{article}
\usepackage{epsfig}
\usepackage{amsmath}
\usepackage{amssymb}
\usepackage{bbm}
\usepackage{verbatim}
\usepackage{bm}
\usepackage{color}
\usepackage{dsfont}
\usepackage[footnotesize]{caption}
\usepackage[subrefformat=parens,labelformat=parens]{subfig}
\usepackage{subfloat}
\usepackage{cancel}
\usepackage{cite}
\numberwithin{equation}{section}

\usepackage{booktabs}
\def\r{\rho}
\def\rb{\bar{\rho}}

\def\Sb{\bar{S}}
\def\p{\phi}
\def\pb{\bar{\phi}}
\def\s{\sigma_0}

\def\d2{W''}
\def\mp{M_{\rm P}}

\begin{document}

\begin{flushleft}
DESY 14-100\\
CPHT-RR037.0614\\
July 2014
\end{flushleft}

\vskip 1cm

\begin{center}
{\Large\bf Large-Field Inflation and Supersymmetry Breaking} 
%\\[2mm]
%  and High-Scale Inflation}

\vskip 2cm

{ Wilfried Buchm\"uller$^a$, Emilian Dudas$^{a,b}$, Lucien Heurtier$^{b,a}$, Clemens Wieck$^a$}\\[3mm]
{\it{
$^a$ Deutsches Elektronen-Synchrotron DESY, 22607 Hamburg, Germany\\
$^b$ CPhT, Ecole Polytechnique, 91128 Palaiseau Cedex, France}
}
\end{center}

\vskip 1cm

\begin{abstract}
\noindent 
Large-field inflation is an interesting and predictive scenario. Its non-trivial embedding in supergravity was intensively studied in the recent literature, whereas its interplay with supersymmetry breaking has been less thoroughly investigated. We consider the minimal viable model of chaotic inflation in supergravity containing a stabilizer field, and add a Polonyi field. Furthermore, we study two possible extensions of the minimal setup. We show that there are various constraints: first of all, it is very hard to couple an O'Raifeartaigh sector with the inflaton sector, the simplest viable option being to couple them only through gravity. Second, even in the simplest model the gravitino mass is bounded from above parametrically by the inflaton mass. Therefore, high-scale supersymmetry breaking is hard to implement in a chaotic inflation setup. As a separate comment we analyze the simplest chaotic inflation construction without a stabilizer field, together with a supersymmetrically stabilized K\"ahler modulus. Without a modulus, the potential of such a model is unbounded from below. We show that a heavy modulus cannot solve this problem.
\end{abstract}

\thispagestyle{empty}

\newpage

%
%%
%%%
%%%%
%%%%%%%%%%%%%%%%%%%%%%%%%%%%%%%%%%%%%%%%%%%%%%%
%%%%%%%%%%%%%%%%%%%%%%%%%%%%%%%%%%%%%%%%%%%%%%%
%%%%%%%%%%%%%%				 %%%%%%%%%%%%%%%%%%%%%%%
%%%%%%%%%%%%%	%         Introduction    %%%%%%%%%%%%%%%%%%%%%%%
%%%%%%%%%%%%%%				  %%%%%%%%%%%%%%%%%%%%%%
%%%%%%%%%%%%%%%%%%%%%%%%%%%%%%%%%%%%%%%%%%%%%%%
%%%%%%%%%%%%%%%%%%%%%%%%%%%%%%%%%%%%%%%%%%%%%%%

\section{Introduction}

Large-field chaotic inflation is an attractive scenario for explaining the initial conditions of the early universe \cite{Linde:1983gd}. In addition to primordial curvature perturbations, which have been measured with remarkable accuracy \cite{Ade:2013uln}, it predicts sizeable tensor perturbations \cite{Lyth:1996im} for which evidence has been reported recently \cite{Ade:2014xna}.
 
The simplest realization of large-field inflation is achieved with a free massive scalar field,
\begin{align}\label{m2inf}
V = m^2 \varphi^2 \, .
\end{align} 
The required flatness of the inflaton potential then requires trans-Planckian field values, $\varphi = \mathcal{O}(10\, \mp)$.  Inflation can nevertheless be treated classically as long as the energy density of the inflaton field does not exceed $\mathcal{O}(\mp^4)$. This is indeed the case for the small inflaton mass $m \sim 10^{-5} \mp$ that is inferred from the measured amplitude of curvature perturbations.

For trans-Planckian field values the contributions of Planck-scale suppressed higher-\linebreak dimensional operators to the inflationary potential are generically relevant. It is therefore important to consider large-field inflation in the context of some ultraviolet completion, for which string theory is the leading candidate, described by supergravity in its low-energy limit. However, the simplest supersymmetric extension of the potential Eq.~\eqref{m2inf} defined by the superpotential 
\begin{align}\label{naiv}
W = \frac12 m \p^2\,,
\end{align}
has a well-known problem. For a typical K\"ahler potential, $K = \p\pb + \ldots$, the supergravity scalar potential, in units where the reduced Planck mass is set to 1,
\begin{align}\label{Vsugra}
V = e^K\left(|\partial_\p W + \partial_\p K W|^2 - 3 |W|^2\right)\,,
\end{align}
is far too steep and inflation is impossible. This problem can be circumvented by choosing a K\"ahler potential with shift symmetry \cite{Kawasaki:2000yn},
\begin{align}
K = \frac{1}{2}(\p+\pb)^2 + \dots \,.
\end{align}
The invariance with respect to $\p \rightarrow \p + i c$, where $c$ is a real constant, implies that $e^{K}$ and $\partial_\p K$ in Eq.~\eqref{Vsugra} are independent of the imaginary part of $\p$, denoted by $\varphi = \sqrt 2\, \text{Im}\,\phi$. Therefore, close to the origin the potential $V(\varphi)$ is now sufficiently flat to allow for inflation. However, the shift symmetry leads to a new problem. Due to the second term in Eq.~\eqref{Vsugra}, for large values of the inflaton field the potential becomes 
\begin{align}
V(\varphi) \sim - 3m^2\varphi^4\,,
\end{align}
and the potential is unbounded from below.

All these problems are avoided by introducing an additional `stabilizer field' $S$, which has no shift symmetry in the K\"ahler potential \cite{Kawasaki:2000yn}, i.e.,
\begin{align}\label{eq:chaoticK}
K = \frac{1}{2}(\p+\pb)^2 + S\Sb \ ,
\end{align}
together with the superpotential
\begin{align}\label{wbilin}
W_{ \rm inf} = m S \p \,, 
\end{align}
which breaks the shift symmetry softly. Up to higher-dimensional terms in the K\"ahler potential the model is determined by an R-symmetry: $R(\p) = 0,\ R(S) = 2$, and a $\mathbbm{Z}_2$-symmetry: $(\p,S) \rightarrow \pm (\p,S)$. In the form defined by Eq.~\eqref{eq:chaoticK} and Eq.~\eqref{wbilin} chaotic inflation predicts for the scalar spectral index $n_s$ and the tensor-to-scalar ratio $r$,
\begin{align}
n_s \simeq 0.967\,, \qquad r \simeq 0.13\,,
\end{align}
for 60 e-folds of inflation. The value of the inflaton field at horizon exit is $\varphi_\star \simeq 15$, and the Hubble scale during inflation is approximately
\begin{align}
H \sim m \varphi_\star \sim 10^{14} \, \text{GeV}\,.
\end{align}

This model has been generalized to a class of chaotic inflation models by replacing the inflaton field $\p$ with a function $f(\p)$ in the superpotential \cite{Kallosh:2010xz}. For recent studies of chaotic inflation in supergravity and string theory, see \cite{Harigaya:2014sua,Harigaya:2014qza,Ferrara:2014ima,Ellis:2014rxa,Li:2014owa,Kallosh:2014xwa,Ketov:2014qha} and \cite{Ibanez:2014zsa,Palti:2014kza,Marchesano:2014mla,Hebecker:2014eua,Arends:2014qca
}, respectively. 

In the following we study chaotic inflation with a stabilizer field together with a sector of supersymmetry breaking, represented by a Polonyi model or an O'Raifeartaigh model. We analyze how different couplings between the two sectors affect the allowed supersymmetry breaking scale and derive upper bounds on the gravitino mass from the requirement of successful inflation. In particular, in none of our setups the gravitino mass can be larger than the Hubble scale during inflation. Among other things, this implies that chaotic inflation is challenged when combined with KKLT moduli stabilization \cite{Kachru:2003aw}, where usually $m_{3/2} > H$ is required \cite{Kallosh:2004yh}. Finally, for the model without a stabilizer field, we consider the possibility that the negative term in the potential Eq.~\eqref{Vsugra} is canceled by adding a supersymmetrically stabilized modulus to the theory. As we shall see, this is not sufficient to obtain a scalar potential bounded from below, contrary to naive expectation.

%
%%
%%%
%%%%
%%%%%%%%%%%%%%%%%%%%%%%%%%%%%%%%%%%%%%%%%%%%%%%
%%%%%%%%%%%%%%%%%%%%%%%%%%%%%%%%%%%%%%%%%%%%%%%
%%%%%%%%%%%%%%				 %%%%%%%%%%%%%%%%%%%%%%%
%%%%%%%%%%%%%	%         Section 2       %%%%%%%%%%%%%%%%%%%%%%%
%%%%%%%%%%%%%%				  %%%%%%%%%%%%%%%%%%%%%%
%%%%%%%%%%%%%%%%%%%%%%%%%%%%%%%%%%%%%%%%%%%%%%%
%%%%%%%%%%%%%%%%%%%%%%%%%%%%%%%%%%%%%%%%%%%%%%%

\section{Chaotic Inflation and Supersymmetry Breaking}

Although chaotic inflation and many of its variants have been extensively studied in the literature, its connection to supersymmetry breaking was not as closely investigated. Generic setups to achieve F-term supersymmetry breaking are the O'Raifeartaigh model \cite{O'Raifeartaigh:1975pr} and the Polonyi model \cite{Polonyi:1977pj}. Coupling the inflaton sector to a supersymmetry breaking sector turns out to be more difficult than expected. In the simplest working scenarios we find that the gravitino mass is bounded from above.

%
%%
%%%
%%%%
%%%%%%%%%%%%%%%%%%%%%%%%%%%%%%%%%%%%%%%%%%%%%%%
%%%%%%%%%%%%%%%%%%%%%%%%%%%%%%%%%%%%%%%%%%%%%%%

\subsection{Minimal chaotic inflation with a Polonyi field}\label{section:minimal}

A minimal way to implement supersymmetry breaking after chaotic inflation is specified by the superpotential\footnote{Notice that this form of superpotential has been studied, in slightly different contexts, in \cite{Kallosh:2011qk,Nakayama:2014xca}.}
\begin{align}\label{eq:Wsimple}
W = m S \phi + f X + W_0 \,, 
\end{align}
i.e., by adding a Polonyi-like sector with a chiral superfield $X$ to the inflation model. Thus, the two sectors decouple except for gravitational-strength interactions. By field redefinitions and a K\"ahler transformation $m$, $f$, and $W_0$ can be chosen to be real; they have mass dimension one, two, and three, respectively. Similar to the superpotential, a suitable K\"ahler potential is obtained by adding the contributions from the inflation and supersymmetry breaking sectors, i.e., 
\begin{align}\label{eq:Ksimple}
K = \frac12(\phi+\bar \phi)^2 + S \bar S  + X \bar X - \xi_1 (X \bar X)^2 - \xi_2 (S \bar S )^2\,.
\end{align}
In the absence of the term proportional to $\xi_2$, the stabilizing scalar  $S$ gets no Hubble-scale
contributions to its mass. This is actually a more general result and applies to any model with or without supersymmetry breaking and a Lagrangian of the type
\begin{align}
K &= K_{a \bar b} \chi_a {\bar \chi}_{\bar b} + S \bar S + K(\phi + {\bar \phi}) + \dots\,, \nonumber \\
W &= W_0 + W_1 (\chi_a,S) + m S \phi \ , \label{m1}
\end{align}
where $\dots$ denotes higher-order terms in the fields $\chi_a$, and $K(\phi + {\bar \phi})$ has at least a quadratic term in an expansion of its argument. On the other hand, the quartic term in the supersymmetry breaking field $X$ is needed to stabilize the corresponding scalar in the true vacuum and circumvent the Polonyi problem.  Thus, the terms proportional to $\xi_1$ and $\xi_2$ are necessary to ensure stability of all directions during inflation and in the ground state.  Both terms may result from integrating out heavy degrees of freedom at the quantum level. Since $X$ gets a Hubble-scale mass during inflation, we often neglect the term involving $\xi_1$ in our discussion of inflation. 

%%%%%%%%%%%%%%%%%%%%%%%
\paragraph{Vacuum after inflation}
$\,$\\
In this combined model, if $f$ is small compared to all other scales in the theory, $m$ still corresponds to the inflaton mass, $f$ denotes the scale of supersymmetry breaking after inflation, and $W_0$ is chosen such that the vacuum energy vanishes after inflation. The latter is achieved if 
\begin{align}
W_0 \simeq \frac{f}{\sqrt3}\,,
\end{align}
at leading order in $f$. After inflation the vacuum of the system is found to lie at
\begin{align}
\langle \phi \rangle = \langle S \rangle = 0\,, \qquad \langle X \rangle \simeq \frac{1}{2 \sqrt3 \xi_1}\,.
\end{align}
In this vacuum the gravitino mass is given by 
\begin{align}
m_{3/2} \simeq W_0 \simeq \frac{f}{\sqrt 3}\,.
\end{align}

However, as will become clear in what follows, this vacuum structure is altered if $f$ is chosen to be larger than $m$. Starting from the full scalar potential
\begin{align}\label{eq:fullPotential}
V = e^K \left\{ | m S + (\phi +\bar \phi) W|^2 + K_{S \bar S}^{-1} |m \phi + K_S W|^2 + K_{X \bar X}^{-1} |f+ K_X W|^2 - 3 |W|^2 \right\}  \,,
\end{align}
with
\begin{align}
K_X &= \bar X (1- 2 \xi_1 |X|^2)\,, \quad K_{XÊ\bar X} = 1- 4 \xi_1 |X|^2\,, \\
K_S &= \bar S (1- 2 \xi_2 |S|^2)\,, \quad \; \; \, \, K_{SÊ\bar S} = 1- 4 \xi_2 |S|^2\,,
\end{align}
we expand $V$ up to second order in all real scalars and obtain
\begin{align}\label{eq:VOrder2}
V = \ &f^2 -3 W_0^2 - 2 \sqrt2 f W_0\, \alpha +2 m W_0 \, \varphi \chi  + \frac12 f^2 \left( 2 \zeta^2 + \chi^2 + \psi^2 \right) \nonumber \\
& -W_0^2 \left( \alpha^2+ \beta^2 + \zeta^2 + \chi^2 + \psi^2 \right) + \frac12 m^2 \left( \zeta^2 + \chi^2  + \psi^2 + \varphi^2  \right) \nonumber \\
 &+ 2 f^2 \xi_1 \left( \alpha^2 + \beta^2 \right)\,,
\end{align}
with
\begin{align}
 S = \frac{\psi +i \chi}{\sqrt{2}}\,, \quad X = \frac{\alpha +i \beta}{\sqrt{2}}\,, \quad \phi = \frac{\zeta +i \varphi}{\sqrt{2}}\,.
\end{align}
From the mass matrix of this system it is evident that assuming $W_0 = \frac{f}{\sqrt3}$ leads to a tachyonic direction close to the origin of the potential if 
\begin{align}
f > m\,.
\end{align}
Specifically, only for $f < m$ there is a stable vacuum at $\langle \phi \rangle = \langle S \rangle = 0$ and $f^2 = 3 W_0^2$ cancels the cosmological constant. For larger $f$ a linear combination of $\phi$ and $S$ obtains a vev and cosmological constant cancellation is ensured by
\begin{align}\label{eq:newW0}
\langle V \rangle = f^2 - 3 W_0^2 + \frac{m^2 \left( f^2 - 6 W_0^2 \right)}{256 \left( f^2 -2 W_0^2 \right)^4 \left( f^2- W_0^2 + 2 \xi_2 \left( f^2 - 3 W_0^2 \right) \right)} = 0\,,
\end{align}
at leading order in $m$. This effect, although small, is taken into account in our analysis of the inflaton dynamics.

%%%%%%%%%%%%%%%%%%%%%%
\paragraph{Interaction during inflation}
$\,$\\
During inflation all fields in the combined system defined by Eq.~\eqref{eq:Wsimple} and Eq.~\eqref{eq:Ksimple} must be stabilized with large masses, i.e., masses larger than the Hubble scale during inflation, so they can be integrated out. Considering the scalar potential in Eq.~\eqref{eq:fullPotential} it is evident that all real scalar degrees of freedom are stabilized at the origin with large masses, except for the inflaton $\varphi = \sqrt 2\, \text{Im}\,\phi$ and the imaginary part of the stabilizer field $\chi = \sqrt 2\, \text{Im}\,S$. Due to the presence of the additional scale $f$ and the constant $W_0$, $\chi$ is shifted from its original minimum at $\langle \chi \rangle = 0$. Assuming that $\chi \ll 1$, we can expand the potential Eq.~\eqref{eq:fullPotential} up to second order around $\chi = 0$. The result reads
\begin{align}
V = f^2 - 3 W_0^2 + \frac12 m^2 \varphi^2 + 2 m W_0 \varphi \chi  + \frac{1}{2} \left(f^2 - 2 W_0^2 +  m^2 + 2m^2 \varphi^2 \xi_2 \right) \chi^2 \,,
\end{align}
neglecting the non-zero vev of $X$. Minimizing this expression with respect to $\chi$ we find 
\begin{align}\label{eq:shiftS}
\chi \simeq -\frac{2m W_0 \varphi}{f^2 - 2 W_0^2 + m^2 + 2 m^2 \varphi^2  \xi_2}\,,
\end{align}
during inflation. Notice that Eq.~\eqref{eq:shiftS} depends on $f$ and $W_0$, as well as on $\varphi$, and that only the imaginary part of $S$ receives a shift. Using a numerical analysis we can verify that $S$ indeed remains stabilized in its new minimum for the entire inflationary epoch. While the inflaton slowly rolls down its quadratic potential the stabilizer field trails its inflaton-dependent minimum near-instantly, see Fig.~\ref{fig:Sstabilized}. 

%%%
\begin{figure}[t]
\centering
      	\subfloat[]{
	\begin{minipage}{0.4\textwidth}\vspace{-5cm}
                \includegraphics[width=1\textwidth]{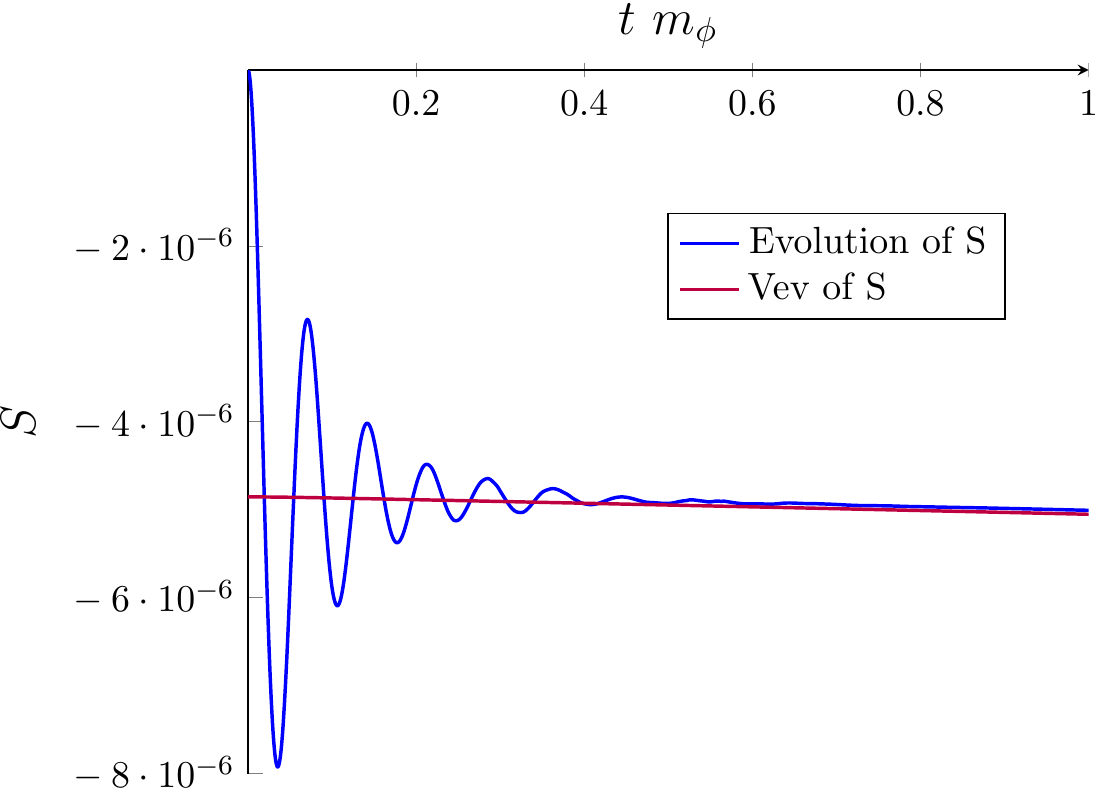}
          \end{minipage}      
          \label{fig:Sstabilized}
 }\hspace{1cm}
  \subfloat[]{
                \includegraphics[width=0.4\textwidth]{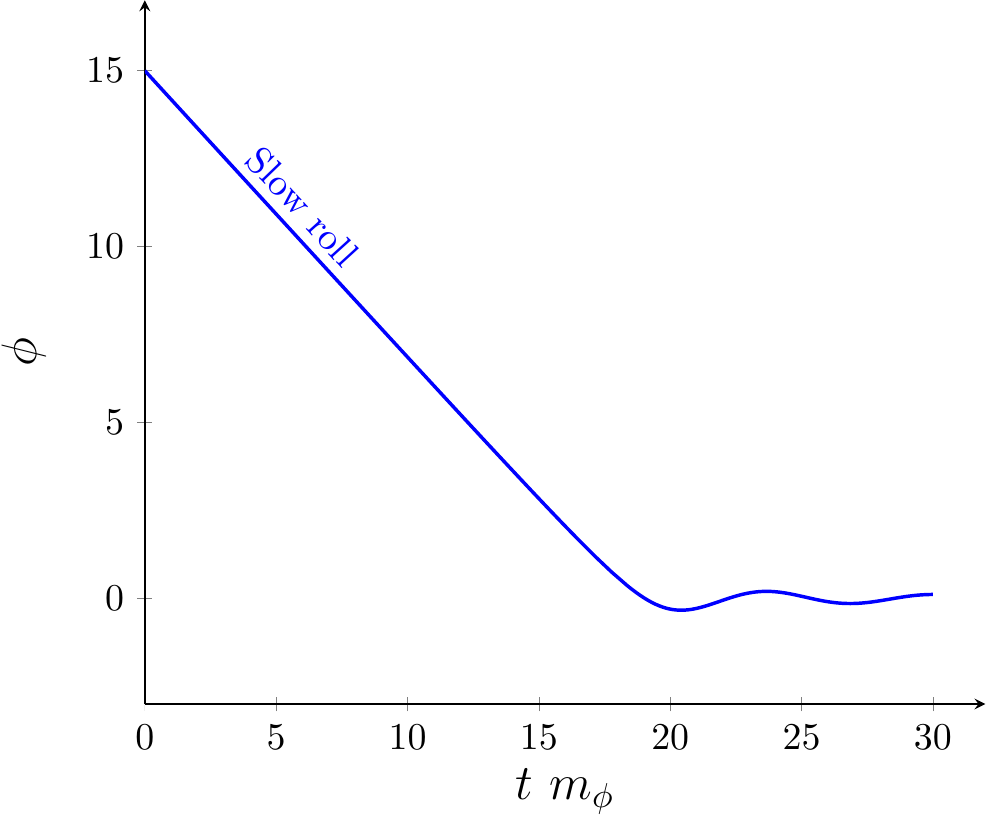}
                \label{fig:infEvolution}
 } 
 \caption{Evolution of the canonically normalized imaginary part of $S$ (a) and the inflaton $\varphi$ (b) during inflation, for $\xi_1 = \xi_2 = 10$ and $f = 10^{-8}$. In this case, since $f < m$, cancellation of the cosmological constant implies $W_0^2 = \frac{f^2}{3}$. Depending on its initial value the stabilizer field settles in its shifted minimum very early and remains stabilized for the rest of the inflationary epoch and beyond (notice the different time scales in the plots). Due to its inflaton-dependence, the vev of $S$ evolves with time.\label{fig:Sstabilized}}
       \end{figure}
%%%
Therefore, $S$ can still be treated as a heavy degree of freedom and can be integrated out at its shifted vev given by Eq.~\eqref{eq:shiftS}. This yields an effective potential for the inflaton direction which reads
\begin{align}\label{eq:effPot}
V(\varphi) = f^2 - 3 W_0^2 + \frac12 m^2 \varphi^2 \left( 1- \frac{4 W_0^2}{f^2 - 2W_0^2 +  m^2 + 2 m^2 \varphi^2 \xi_2} \right)\,.
\end{align}
Evidently, depending on the magnitude of $f$ and hence the gravitino mass, the correction resulting from integrating out $S$ may severely alter the predictions of chaotic inflation.

%%%%%%%%%%%%%%%%%%%%%%
\paragraph{Bounds on the gravitino mass}
$\,$\\
Considering the effective inflaton potential Eq.~\eqref{eq:effPot}, alteration of the CMB observables, in particular the scalar spectral index $n_s$ and the tensor-to-scalar ratio $r$, is to be expected at $f \gtrsim m$. We expect that increasing $f$ even further will make inflation unfeasible at a value which satisfies
\begin{align}\label{}
3 m^2 \lesssim f^2 \lesssim 2 m^2 \varphi^2 \xi_2\,,
\end{align}
neglecting the correction to $W_0$ in Eq.~\eqref{eq:newW0}. Since $m$ is fixed by observations to be ${m\simeq6\times10^{-6}}$ in Planck units, it is necessary to specify realistic values of $\xi_2$ to obtain a meaningful upper bound on the gravitino mass. We assume that the K\"ahler potential terms involving $\xi_1$ and $\xi_2$ stem from couplings of heavy modes to $S$ and $X$, i.e., from
\begin{align}
W_\text{heavy} \supset \lambda_1 S \psi_1^2 + \lambda_2 X \psi_2^2 +\text{mass terms}\,,
\end{align}
where $\psi_i$ denotes heavy modes of mass $M_i$. Then a quartic term for $S$ in $K$ is generated by one-loop quantum corrections of the Coleman-Weinberg type,
\begin{align}\label{eq:quantumKahler}
K_\text{1-loop} \simeq S \bar S \left[ 1- \frac{\lambda^2}{16 \pi^2} \log{\left( 1+ \frac{\lambda^2 S \bar S}{M^2} \right)} \right] \simeq S \bar S - \frac{\lambda^4}{16 \pi^2 M^2} (S \bar S)^2\,,
\end{align}
and similarly for $X$, cf.~the discussion in \cite{Kallosh:2006dv}. Thus, in the generic case $\lambda_i \sim \mathcal O(1)$ the $\xi_i$ are related to the mass scales $M_i$ as follows,
\begin{align}\label{eq:xi}
\xi_i \sim \frac{1}{16 \pi^2 M_i^2}\,.
\end{align}
Since the heavy degrees of freedom should be integrated out above the energy scale during inflation, $M_i \gtrsim \rho_\text{inf} \sim M_\text{GUT} \simeq 0.01$, but below the Planck scale, we assume 
\begin{align}
\xi_1 \simeq \xi_2 \simeq 10
\end{align}
to be reasonable values for the coefficients.\footnote{Note that quartic terms in the K\"ahler potential could also arise from $\alpha'$ corrections in string theory. In such a setup the coefficients would rather be $\xi \sim \frac{1}{M_\text{s}^2}$, where $M_\text{s}$ denotes the string scale. In order for string modes to decouple, $M_\text{s}$ would have to be larger than the energy scale during inflation, but smaller than the Planck scale. Due to the absence of the loop suppression factor $16 \pi^2$, this could result in substantially larger coefficients.} 

\begin{figure}[t]
\centering
      	\subfloat[]{
	\begin{minipage}{0.4\textwidth}\vspace{-5cm}
                \includegraphics[width=1\textwidth]{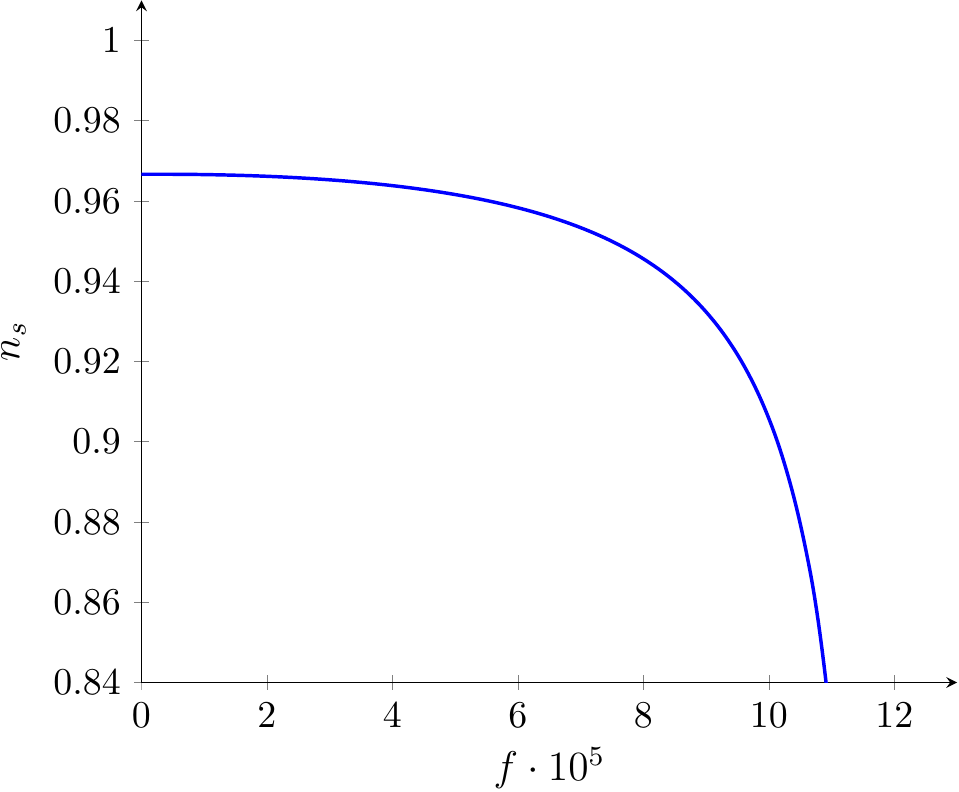}
          \end{minipage}      
          \label{fig:nsMinimal}
 }\hspace{1cm}
  \subfloat[]{
                \includegraphics[width=0.4\textwidth]{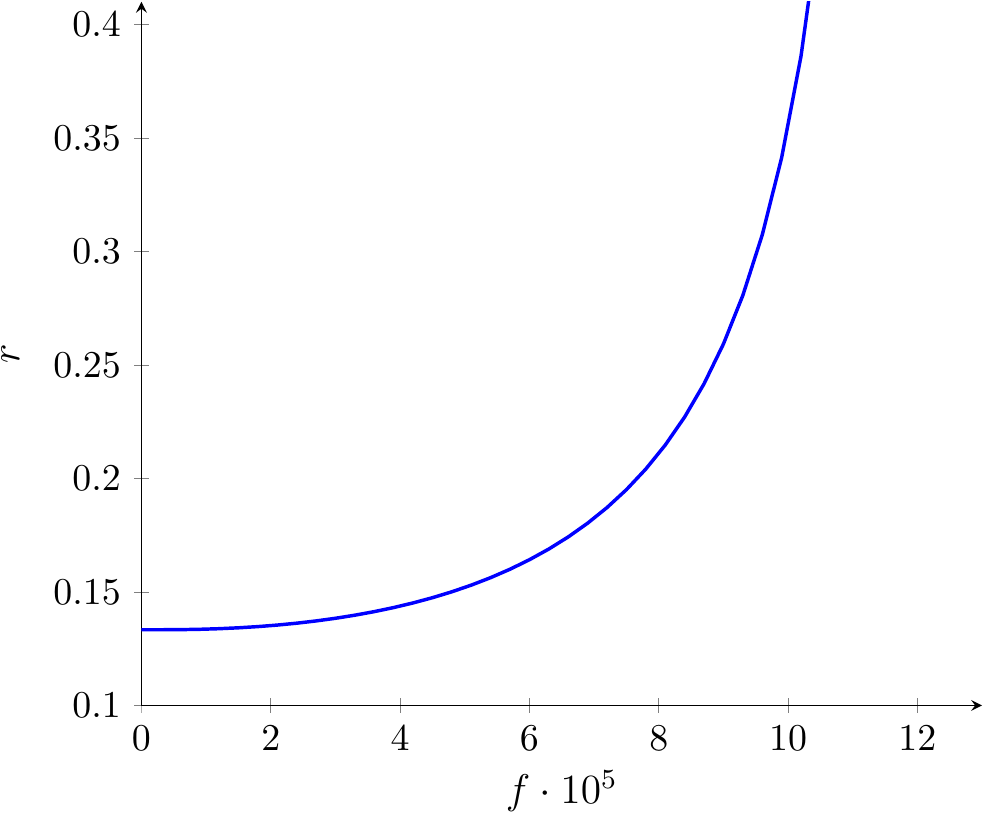}
                \label{fig:rMinimal}
 } 
 \caption{$n_s$ and $r$ as a function of the supersymmetry breaking scale $f$. Clearly, the model is ruled out by observation at values of $f$ quite below $10^{-4}$.\label{fig:nsandr}}
       \end{figure}

Using these estimates, the spectral index and tensor-to-scalar ratio resulting from the corrected potential Eq.~\eqref{eq:effPot} are displayed in Fig.~\ref{fig:nsandr}, as a function of $f$.
%%%
%%%
Evidently, above a value of $f \simeq 8 \times 10^{-5}$ the tensor-to-scalar ratio increases above 0.2 and $n_s$ drops below 0.94, a point at which the model is essentially ruled out by observation. This translates into a bound on the gravitino mass,
\begin{align}\label{eq:obsBound}
m_{3/2} \lesssim 10^{14} \, \text{GeV}\,.
\end{align}
Therefore, the most minimal way of achieving supersymmetry breaking in chaotic inflation excludes the possibility $m_{3/2} \gtrsim H$. This may have interesting implications for setups with string-inspired supersymmetry breaking in which the supersymmetry breaking scale is usually very high, as recently investigated in \cite{Ibanez:2014zsa, Palti:2014kza,Hebecker:2014eua}.

%
%%
%%%
%%%%
%%%%%%%%%%%%%%%%%%%%%%%%%%%%%%%%%%%%%%%%%%%%%%%
%%%%%%%%%%%%%%%%%%%%%%%%%%%%%%%%%%%%%%%%%%%%%%%

\subsection{Effects of additional interactions}

As an attempt to relax the gravitino mass bound \eqref{eq:obsBound} it is possible to extend the previous minimal model by a coupling between $X$ and $\phi$ in the superpotential which preserves the R-symmetry,
\begin{align}\label{model2}
W = m S \phi + M X \phi + f X + W_0 \,.
\end{align}
The new mass scale $M$ contributes, together with $m$, to the mass of the inflaton, i.e.,
\begin{align}
V = \frac12 m^2 \varphi^2 \longrightarrow V = \frac12 (m^2 + M^2) \varphi^2\,,
\end{align}
in the absence of supersymmetry breaking. The associated K\"ahler potential can be written as\begin{align}
K = \frac12(\phi+\bar \phi)^2 + S \bar S  + X \bar X - \xi_1 (X \bar X)^2\,.
\end{align}
Notice that no quartic term in $S$ is needed to stabilize the corresponding scalars in this setup. As before, we can always choose $W_0$, $f$, and $m$ to be real, but the mass $M$ is generically complex. For simplicity, we take it to be real in what follows. 

%%%%%%%%%%%%%%%%%%%%%%
\paragraph{Vacuum after inflation}
$\,$\\
In this framework, the fields are stabilized at different vevs in the vacuum, and the constant $W_0$ consequently takes a different value to cancel the cosmological constant. Specifically, writing the complex scalars in terms of their real components 
\begin{align}
 S = \frac{\psi +i \chi}{\sqrt{2}}\,, \quad X = \frac{\alpha +i \beta}{\sqrt{2}}\,, \quad \phi = \frac{\zeta +i \varphi}{\sqrt{2}}\,,
\end{align}
the associated vacuum expectation values after inflation are given by
\begin{align}
 \langle \varphi \rangle = \langle \chi \rangle = \langle \beta \rangle=0\,, \quad \langle \zeta \rangle\simeq-\sqrt{2}\frac{M f}{m^2+M^2}\,, \quad \langle \alpha \rangle \simeq \frac{ 1 }{\sqrt 6 \, \xi_1} \frac{m}{ \sqrt{m^2 + M^2}}\,,
 \end{align}
 and
 \begin{align}
 \langle \psi \rangle\simeq\frac{M}{\left(m^2+M^2\right)^{3/2}}\frac{f^2 \left(m^2+3 M^2\right)-3 m^2 \left(m^2+M^2\right)}{3 \sqrt{6}~ \xi_1~ m^2}\,,
\end{align}
at leading order in $f$ and $1/\xi_1$. The gravitino mass in the true vacuum is given by
\begin{align}\label{eq:extendedW0}
m_{3/2} \simeq W_0 \simeq \frac{m}{\sqrt{m^2+M^2}}\frac{ f}{\sqrt{3}}\,.
\end{align}
Notice that, as in the model discussed in Section \ref{section:minimal}, this vacuum will be corrected for large values of $f$. The corrections will, however, not alter our conclusions about the allowed gravitino mass by much. Therefore, in what follows we use the value of $W_0$ stated in Eq.~\eqref{eq:extendedW0} as a leading-order approximation.

%%%%%%%%%%%%%%%%%%%%%%
\paragraph{Interaction during inflation}
~\newline
As in the decoupled model discussed in Section \ref{section:minimal}, the supersymmetry breaking scale $f$ induces a shift of the imaginary part of $S$ during inflation. In fact, some of the other real scalars are shifted as well, but their vevs are suppressed compared to that of $\chi$ and will therefore be neglected in what follows. A numerical analysis once more confirms that all vevs are reached quickly and that all fields, except the inflaton $\varphi$, remain stabilized during inflation. In the same manner as in the previous section, expanding up to second order in $\chi$ and integrating out the field gives a leading-order effective potential for the inflaton. The result reads
\begin{align}\label{eff2}
V(\varphi)\; =\;Ê\; &\frac12 (1 + \delta^2)m^2 \varphi^2 \left( 1- \frac{8 f^2}{f^2 (2+ 8 \delta^2 + 6\delta^4) + 3 m^2 (1+ \delta^2)^2 (2+ \delta^2 \varphi^2)} \right) \nonumber \\
 &+ f^2\left(1-\frac{1}{1+\delta^2}\right)\,,
\end{align}
where we have introduced the dimensionless parameter 
\begin{align}
\delta = \frac{M}{m}\,.
\end{align}
Notice that, in the limit $\delta \to 0$, Eq.~\eqref{eff2} reduces to the effective potential of the minimal model in Section \ref{section:minimal}, given by Eq.~\eqref{eq:effPot}. The only difference is that in the present setup $\xi_2 = 0$. Again, it appears that in this model chaotic inflation is not possible for arbitrarily large values of $f$.

%%%%%%%%%%%%%%%%%%%%%%
\paragraph{Bounds on the gravitino mass}~\newline
In fact, it turns out there is a more stringent upper bound on the gravitino mass. The region of parameter space where the model reduces to a single-field inflation model is $\delta > \mathcal O(1)$. In this case, there is again a bound stemming from the alteration of the CMB observables due to the presence of $f$.

\begin{figure}\centering
\begin{tabular}{c c}
\includegraphics[width=0.46\textwidth]{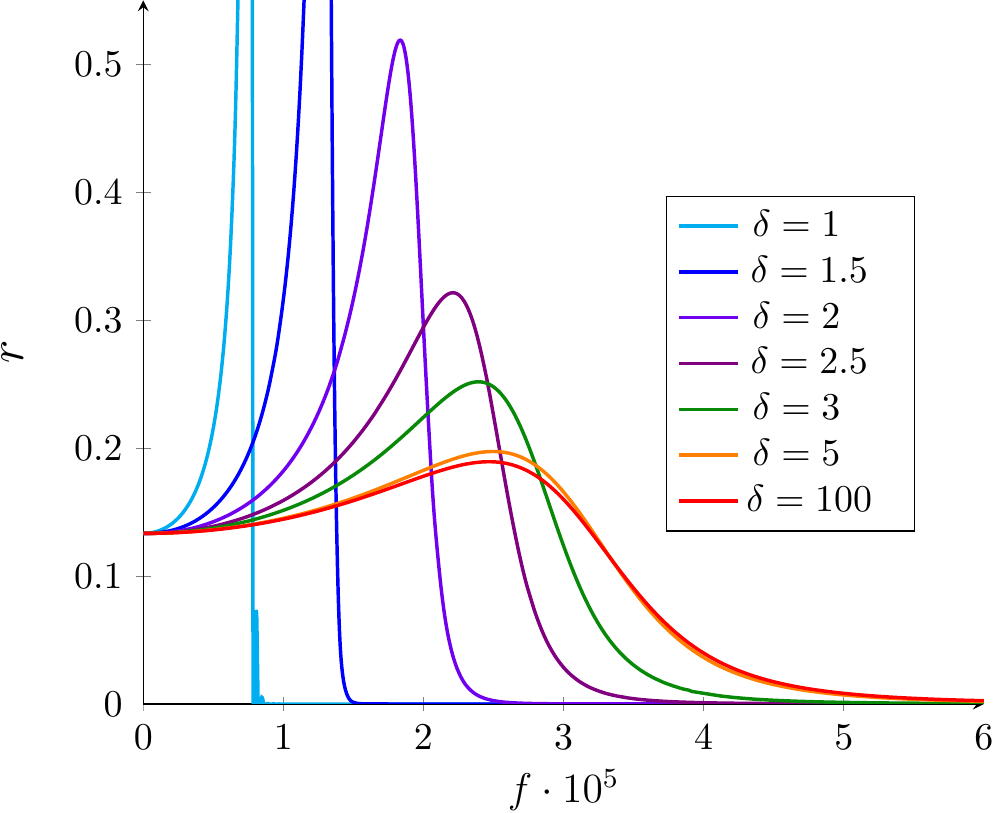}&  \includegraphics[width=0.47\textwidth]{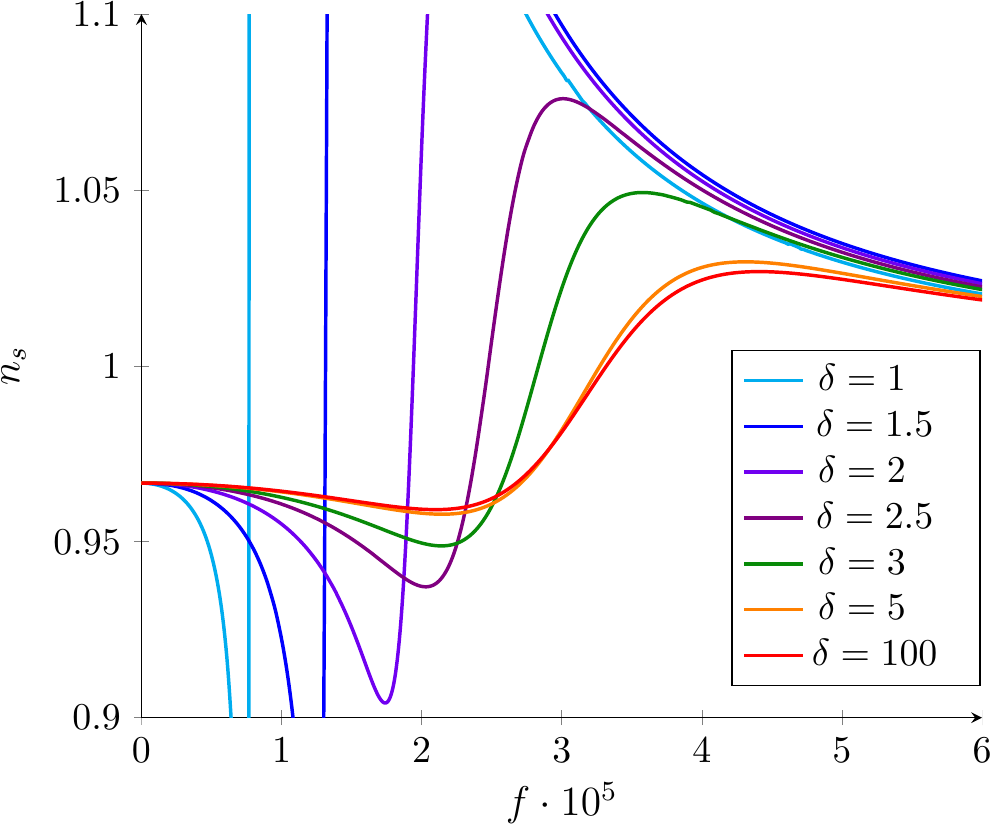}\\
\end{tabular}
\caption{\label{fig:obsbound}{\footnotesize CMB observables as functions of the supersymmetry breaking scale $f$, for different values of $\delta$ and $m=6\times 10^{-6}$, $\xi_1=10$ in Planck units.}}
\end{figure}

To visualize this bound, and how it scales with $\delta$, the observables $n_s$ and $r$ are depicted in   
Fig.~\ref{fig:obsbound} as functions of $f$.
For small values of $\delta$, there is a bound from demanding that $r$ does not surpass 0.2 and $n_s$ does not drop below 0.94, analogous to Section \ref{section:minimal}. For larger values of $\delta$, however, these requirements are always fulfilled and a bound on $f$ arises from demanding that $n_s$ does not surpass $\sim 0.98$. Although increasing $\delta$ pushes the bound to slightly higher values of $f$, this effect saturates at roughly $\delta \sim 10$. However, since for $\delta \gg 1$ the gravitino mass can be written as 
\begin{align}
 m_{3/2} =  \frac{1}{\sqrt{1+\delta^2}}\frac{ f}{\sqrt{3}}\simeq  \frac{1}{\delta}\frac{ f}{\sqrt{3}}\,,
\end{align}
increasing $\delta$ will, at a certain point, actually make the upper bound on $m_{3/2}$ more stringent. It turns out that the least severe upper bound on $m_{3/2}$ is obtained for $\delta \simeq 4$, in which case 
\begin{align}
f \lesssim 3 \times 10^{-5} \quad \Rightarrow \quad m_{3/2} \lesssim 8 \times 10^{12}\, \text{GeV}\,.
\end{align}
Clearly the attempt to relax the bound obtained in the decoupled model of Section \ref{section:minimal} was not successful, since now $m_{3/2} \lesssim 0.1\, H$. One may suspect that this is due to the absence of the large stabilizing term proportional to $\xi_2$. Indeed, including this term in the present setup trivially reproduces the mass bound Eq.~\eqref{eq:obsBound} in the limit $\delta \to 0$. Whenever $M$, and thus $\delta$, is non-zero, however, the additional coupling will make the bound more severe. In particular, in the regime $\delta \sim \mathcal O(1)$ the upper bound on $m_{3/2}$ is very close to the bound obtained using $\xi_2 = 0$.

%
%%
%%%
%%%%
%%%%%%%%%%%%%%%%%%%%%%%%%%%%%%%%%%%%%%%%%%%%%%%
%%%%%%%%%%%%%%%%%%%%%%%%%%%%%%%%%%%%%%%%%%%%%%%

\subsection{Supersymmetry breaking in the O'Raifeartaigh model}

A minimal way to incorporate chaotic inflation and supersymmetry breaking seems to be contained in the O'Raifeartaigh model, without the addition of extra fields or couplings. In particular, writing the superpotential of \cite{O'Raifeartaigh:1975pr} as 
\begin{align}
W = X (f + \frac{1}{2} h S^2 ) + m S \phi  + W_0 \,,
\end{align}
where the stabilizer $S$ and the inflaton $\phi$ take up the roles of the two - usually heavy - ``O'Raifeartons" and the F-term of $X$ breaks supersymmetry. If $\phi$ is protected by a shift symmetry, as in the cases studied before, the tree-level K\"ahler potential takes the form
\begin{align}
K = \frac{1}{2} (\phi + \bar \phi)^2 + S \bar S + X \bar X\,.
\end{align}
Again we can choose $W_0$, $f$, and $m$ to be real, in which case the Yukawa coupling $h$ can be complex. For simplicity, we take it to be real in what follows. In this setup, inflation should be possible in the direction of the imaginary part of $\phi$. After inflation, $\phi$ is stabilized at the origin and supersymmetry is broken by $X$ as in the Polonyi model.

However, upon closer inspection the model turns out to be problematic due to tachyonic instabilities during inflation. The F-term of $S$ induces contributions in the scalar potential of the form
\begin{align}
V \supset m\varphi \, X \bar S + \text{c.c.}\,,
\end{align}
i.e., there are mass eigenstates with squared mass 
\begin{align}
m_\text{tach}^2 \sim - m \varphi \sim - HÊ\,.
\end{align}
Considering the original O'Raifeartaigh model and the discussion involving Eq.~\eqref{eq:quantumKahler} one may hope that quantum corrections from integrating out $S$ can lift these tachyonic directions, but they can not. In order to induce a loop-generated mass term of a size comparable to $\sqrt H \sim M_\text{GUT}$, heavy modes would have to be integrated out far below the GUT scale. In other words, the coefficient $\xi$ in Eq.~\eqref{eq:xi} would have to be larger than allowed by the effective field theory if the new states are heavy enough to not perturb the single-field inflation dynamics.

There are two notable ways out to make an O'Raifeartaigh model interacting with the inflaton viable. The first one is invoking microscopic (string theory) contributions to the K\"ahler potential of the form $(1/\Lambda_\text{UV}^2) |S|^4$, with a UV cut-off $\Lambda_\text{UV} \lesssim M_\text{GUT}$. During inflation, these would generate large mass terms for $S$ which would cure the tachyonic contributions. A string theory with $M_\text{s} \sim M_\text{GUT}$ and $\alpha'$-corrections to the K\"ahler potential, plus some additional assumptions on the origin of $S$, could be responsible for the existence of such terms. 

A second solution could be to add a term of the type $\xi_1 |X|^4$ with a very large coefficient $\xi_1$. This would decouple the sgoldstino scalar and again could cure the problem. Technically, this is equivalent to working with a constrained goldstino superfield $X^2=0$. The solution is
\begin{align}
X = {\psi_X \psi_X \over 2 F_X} + \sqrt{2} \theta \psi_X + \theta^2 F_X \,, \label{or1}
\end{align}
and  leads to a non-linearly realized supersymmetry which was discussed in other inflationary contexts in \cite{AlvarezGaume:2010rt,AlvarezGaume:2011db,Antoniadis:2014oya,Ferrara:2014ima,Ferrara:2014fqa}. In our case, the effective action is described by  
\begin{align}
K &= \frac{1}{2} (\phi + \bar \phi)^2 + S \bar S + X \bar X - \xi |S|^4\, , \nonumber \\
W &= X \left(f + \frac{1}{2} h S^2 \right) + m S \phi  + W_0\,,  \quad {\rm and } \quad X^2 = 0 \, . \label{or2}
\end{align}
Since the superfield $X$ contains no scalar, the dynamics simplifies. As in the previous examples, the only relevant fields during inflation are the inflaton $\varphi$ and ${\rm Im} S = \chi/\sqrt{2}$. At all orders in the inflaton and quadratic order in $S$, the scalar potential is
\begin{align}
V \simeq f^2 - 3 W_0^2 + \frac{1}{2}m^2 \varphi^2 + 2 m W_0 \varphi \chi+ \frac{1}{2} \left( f^2 - 2 W_0^2 - hf + m^2 + 2 m^2 \varphi^2 \xi \right) \chi^2
  \, . \label{or3}
\end{align}
As in the simpler models before, the field $\chi$ will track the inflaton trajectory, with a value given by
\begin{align}
\chi = - \frac{2 m W_0 \varphi}{f^2- 2 W_0^2 - hf + m^2 + 2m^2 \varphi^2 \xi} \, . \label{or4}
\end{align}
Since the tracking is very fast, we can write down an effective inflaton potential by inserting Eq.~\eqref{or4} into the scalar potential Eq.~\eqref{or3}. The result reads
\begin{align}
V (\varphi) = f^2 - 3 W_0^2 +\frac12 m^2 \varphi^2 \left( 1 - \frac{4 W_0^2}{f^2- 2 W_0^2 - hf + m^2  + 2 \xi m^2 \varphi^2} \right)\, . \label{or5}
\end{align}
Notice that, as in Section 2.2, we neglect the sub-leading correction stemming from the modified cosmological constant cancellation condition for large $f$. 

As is well-known, the O'Raifeartaigh model has two vacua, depending on the values of the parameters:
\begin{itemize}

\item $|h f| > m^2$ \\
In this case, either the imaginary or the real part of $S$ has a non-zero vev in the ground state in the rigid supersymmetric limit, equal to $ \sqrt{2 (|hf|-m^2)}/h$. Cancellation of the cosmological constant at leading order is in this case ensured by
\begin{align}
m^2 (2 |hf|-m^2) \simeq 3 h^2 |W_0|^2  \, . \label{or7}
\end{align}  
The gravitino mass in the ground state is given by
\begin{align}
m_{3/2} \simeq \frac{m}{\sqrt{3}h} \sqrt{2 |hf|- m^2} \,,
\end{align}
which, for $h \sim O(1)$, is bounded by 
\begin{align}\label{eq:m32smallerm}
m_{3/2} < m \,.
\end{align}
However, even if $h$ is chosen to be very small to avoid this bound, we expect the CMB observables to receive similar corrections as in the model discussed in Section 2.1, as soon as $f \gtrsim m$.

\item $|h f| < m^2$ \\
In this case, all fields are stabilized at the origin in the true vacuum and $W_0 = \frac{f}{\sqrt3}$ cancels the cosmological constant. Again, as can be deduced from the similarity of the effective inflaton potentials, the analysis of Section 2.1 applies to good approximation. Therefore, we expect a bound on the gravitino mass of 
\begin{align}
m_{3/2} \lesssim H\,.
\end{align}

\end{itemize}

In summary, in the O'Raifeartaigh model with non-linear supersymmetry, imposed by the constraint $X^2=0$, the outcome is again an upper bound on the gravitino mass which is similar to the ones obtained in the previously discussed models. 

We have proposed two solutions to make an O'Raifeartaigh model coupled non-trivially to the inflaton
viable from the chaotic inflation perspective. The simplest option, however, is clearly to decouple
the supersymmetry breaking sector containing the fields $\chi_i$ from the inflaton sector containing
the inflaton $\phi$ and the stabilizer $S$, like for example in models with a superpotential
\begin{align}
W = W_\text{O'R} (\chi_i) + m S \phi \ . \label{or9}
\end{align}
The model of Section \ref{section:minimal} is probably the simplest example of this type. 

%%%
%%%%
%%%%%%%%%%%%%%%%%%%%%%%%%%%%%%%%%%%%%%%%%%%%%%%
%%%%%%%%%%%%%%%%%%%%%%%%%%%%%%%%%%%%%%%%%%%%%%%
%%%%%%%%%%%%%%				 %%%%%%%%%%%%%%%%%%%%%%%
%%%%%%%%%%%%%	%         Section 3       %%%%%%%%%%%%%%%%%%%%%%%
%%%%%%%%%%%%%%				  %%%%%%%%%%%%%%%%%%%%%%
%%%%%%%%%%%%%%%%%%%%%%%%%%%%%%%%%%%%%%%%%%%%%%%
%%%%%%%%%%%%%%%%%%%%%%%%%%%%%%%%%%%%%%%%%%%%%%%

\section{Chaotic Inflaton with Stabilized Moduli}

As a separate point, we study in the following the simplest chaotic inflation model in supergravity without the stabilizer field $S$, including the effect of supersymmetry breaking. The simplest model of this type is described by
\begin{align}
K & =  \frac{1}{2}(\p+\pb)^2 + X \bar X - \xi_1 (X \bar X)^2\, , \nonumber \\
W_{ \rm inf}&= \frac{1}{2}m \p^2 + f X + W_0 \, . \label{s1}
\end{align}
However, this model suffers from the same instability problem as in the absence of supersymmetry breaking, as mentioned in the Introduction. This becomes evident by inspecting the scalar potential,
\begin{align}
V = e^K \Bigg\{ 
&\left| m \phi \left[1 + \frac12 \p (\p+\pb)\right]  + (f X + W_0) (\p+\pb) \right|^2 \nonumber \\ 
&+ K_{X \bar X}^{-1}\left|f + {K_X} \left(\frac12 m \p^2 + f X + W_0\right)\right|^2 - 3 \left| \frac{m \p^2}{2} + f X + W_0 \right|^2  \Bigg\}\,.
\label{s2}
\end{align}
Indeed, we recover a potential unbounded from below for large $\p$. Since the origin of the problem is the negative supergravity contribution to the potential, at first sight, the presence of a supersymmetrically stabilized modulus $\rho$ can solve the problem via a no-scale cancellation. Note that the effects of stabilized K\"ahler moduli on similar models of chaotic inflation have been previously studied in \cite{Davis:2008fv,Kallosh:2011qk,Buchmuller:2014vda}. Starting from the K\"ahler potential\footnote{Throughout this paper we consider the simplest viable K\"ahler potentials. In a microscopic setup like string theory they are usually more involved, depending on the precise origin of the inflaton multiplet and its couplings to the heavy stabilized and lighter moduli. For example, if in type IIB orientifolds with D3 and D7 branes the inflaton is the position of a D3 (D7) brane, at lowest order its K\"ahler potential will mix (will not mix) with $T$. While our analysis assumes the absence of mixing, the final result is expected to hold in the presence of mixing as well, provided the modulus is stabilized in a supersymmetric way. A detailed analysis of all these cases would be interesting but is beyond the scope of our paper.} 
\begin{align}
K = -3 \log{(\r + \rb)} + \frac{1}{2}(\p+\pb)^2 + X \bar X  - \xi_1 (X \bar X)^2\,, \label{s3}
\end{align}
and superpotential
\begin{align}
W &= W_{\rm mod}(\r)  + W_{ \rm inf}(\p,X) \, , \\
W_{ \rm inf}(\p,X) &= \frac12 m \p^2 + f X + W_0\, , \label{s4}
\end{align}
the scalar potential reads
\begin{equation}
V = e^K \left\lbrace 
\frac{(\r + \rb)^2}{3} |\partial_\r W|^2 - (\r + \rb) 
(\partial_\r W {\overline W} +  \overline{\partial_\r W} W ) +  K^{\alpha \bar \alpha} D_\alpha W  D_{\bar \alpha} \overline W  \right\rbrace \, ,  \label{s5}
\end{equation}
where $\alpha = \p,X$. Thus, the dangerous negative term indeed seems to be canceled due to the no-scale structure of the model.
%\footnote{Our results are, however, valid for more general K\"ahler potentials.}

However, upon closer inspection this cancellation does not occur. During inflation, there is a non-trivial interaction between the inflaton and the modulus field. The modulus vev is shifted by an amount $\delta \r$, which can be evaluated in an inverse expansion of the modulus mass \cite{Buchmuller:2014vda}, assuming it is heavy enough. Similar setups with heavy K\"ahler moduli have been previously studied in \cite{Kallosh:2004yh,Linde:2011ja,Dudas:2012wi,Kallosh:2014oja}.  We assume that $W_\text{mod}$ is such that the scalar potential has a local minimum at $\r_0 = \rb_0 \equiv \s$ which is supersymmetric and Minkowski,
\begin{align}
D_\r W_{\rm mod}(\s) = W_\text{mod}(\sigma_0) = 0 \, .  \label{s6}
\end{align}
The mass of the modulus in the ground state is given by
\begin{align}
m_\r = \frac{\sqrt{2\s}}{3} \,W_\text{mod}''(\sigma_0)\,,   \label{s7}
\end{align}
where primes denote derivatives with respect to $\rho$. Notice that this stabilization scheme differs from the original proposal by KKLT, in the sense that $m_\rho$ and $m_{3/2}$ are uncorrelated and the modulus can be much heavier than the gravitino.

For $m_{\r} > H$, in order to guarantee single-field inflation, the shift of the modulus vev $\delta \r$ can be expanded in powers of $H/m_\r$ and is given, at leading order, by 
\begin{equation}
\delta \r \simeq \frac{W_\text{inf}}{\sqrt{2 \sigma_0} m_\rho}\, , \label{s8}
\end{equation}
which is small, $\delta \r \leq \sigma_0$, if $m_\r > \frac{W_\text{inf}}{(2 \sigma_0)^{3/2}}$. 

As demonstrated in \cite{Buchmuller:2014vda}, once the modulus is integrated out at its shifted vev the inflaton potential is corrected by terms which can be expanded in powers of $H/m_\rho$. The modified inflaton potential reads 
\begin{align}\label{eq:effpot} 
V= \frac{V_\text{inf}(\phi_\alpha)}{(2 \sigma_0)^3} - \frac{3}{2 (2\s)^{9/2} \,m_\rho} \Big\{ W_\text{inf} \Big[V_\text{inf}(\phi_\alpha) + e^K K^{\alpha \bar \alpha} \partial_\alpha W_\text{inf} D_{\bar \alpha} \overline W_\text{inf} \Big] + \text{ c.c.} \Big\} \,,
\end{align}
at leading order in $H/m_\rho$. Here $V_\text{inf}(\phi_\alpha)$ denotes the inflationary potential in the absence of a modulus sector. But this is precisely the potential before the addition of the modulus, Eq.~(\ref{s2}), which is unbounded from below. The leading order correction in Eq.~\eqref{eq:effpot} may be sizeable, depending on $m_\rho$, but cannot solve the problem of unboundedness from below. Therefore, after integrating out $\rho$ at its true minimum the no-scale cancellation is not effective since the modulus minimizes its F-term, including the contribution from the inflaton sector. 

This is actually to be expected: if the modulus does not break supersymmetry and is heavy enough to not perturb the single-field chaotic inflation, it will decouple from the inflationary dynamics at leading order. Then the leading-order scalar potential can be obtained in the limit $m_\r \to \infty$ which results in the original model defined by Eq.~(\ref{s1}). A lighter modulus can certainly change the situation, but for $m_\rho < H$ the model turns into a much more complicated multi-field inflation model. Another solution to this problem could be to stabilize the modulus non-supersymmetrically. If $\rho$ has a non-vanishing F-term during inflation, it may indeed cancel the dangerous $-3|W|^2$ term by virtue of the no-scale structure. A similar setup has recently been discussed in \cite{Hebecker:2014eua}, where the modulus is stabilized non-supersymmetrically in a Large Volume Scenario.

%
%%
%%%
%%%%
%%%%%%%%%%%%%%%%%%%%%%%%%%%%%%%%%%%%%%%%%%%%%%%
%%%%%%%%%%%%%%%%%%%%%%%%%%%%%%%%%%%%%%%%%%%%%%%
%%%%%%%%%%%%%%				 %%%%%%%%%%%%%%%%%%%%%%%
%%%%%%%%%%%%%	%         Section 4        %%%%%%%%%%%%%%%%%%%%%%%
%%%%%%%%%%%%%%				  %%%%%%%%%%%%%%%%%%%%%%
%%%%%%%%%%%%%%%%%%%%%%%%%%%%%%%%%%%%%%%%%%%%%%%
%%%%%%%%%%%%%%%%%%%%%%%%%%%%%%%%%%%%%%%%%%%%%%%

\section{Conclusion}

The motivation of this work is twofold. On the one hand, the main subject of study was the interplay between large-field inflation and supersymmetry breaking. Our findings reinforce the models with small value of the superpotential during inflation, among which we studied models with a `stabilizer' field coupled to the inflaton and a supersymmetry breaking sector. We found that models with renormalizable couplings of non-gravitational origin between the inflaton and the supersymmetry breaking sector are very constrained and difficult to construct. Therefore, the simplest viable models turn out to be the ones in which the coupling between the two sectors is purely gravitational. In all cases we studied, we found an upper bound on the supersymmetry breaking scale and consequently on the gravitino mass. The precise bound is model-dependent but parametrically of the order of the inflaton mass. Therefore, chaotic inflation is challenged in scenarios like KKLT moduli stabilization, where usually $m_{3/2} > H$ is required. Let us stress that models with ``strong moduli stabilization" \cite{Kallosh:2004yh,Linde:2011ja,Dudas:2012wi,Kallosh:2014oja} with a light gravitino, $m_{3/2} \ll m$, were constructed some time ago and are perfectly viable from our perspective, in particular with regard to low-energy supersymmetry or mini-split models. Our results merely emphasize that the complementary high-mass region $m_{3/2} > m$, interesting in some string constructions, is more problematic for chaotic inflation, even in such models.

On the other hand, removing the stabilizer field (re)introduces the problem that the scalar potential is unbounded from below. We found that coupling the inflaton to a supersymmetrically stabilized modulus does not help in this respect, if the modulus is heavy enough to not perturb the chaotic inflationary dynamics. 

While we do not claim the existence of a no-go theorem, our results imply non-trivial constraints on high-scale supersymmetry breaking scenarios, if the inflationary dynamics is of large-field, chaotic type. Models with moderately small (compared to the inflaton mass) scale of supersymmetry breaking \cite{Kallosh:2004yh,Linde:2011ja,Dudas:2012wi,Kallosh:2014oja} are therefore preferable from this viewpoint. 

\subsection*{Acknowledgments}
The authors thank Arthur Hebecker, Andrei Linde, Alexander Westphal, and Martin Winkler for stimulating discussions. This work has been supported by the German Science Foundation (DFG) within the Collaborative Research Center 676 ``Particles, Strings and the Early Universe'' and  by  the ERC advanced grant "MassTeV".  E.D. and L.H. would like to thank the Alexander von Humboldt foundation and DESY-Hamburg for support and hospitality. The work of C.W. is supported by a scholarship of the Joachim Herz Stiftung.

\end{document}